\documentclass[aps,twocolumn,amsmath,amssymb,prl,superscriptaddress]{revtex4-1}       

\usepackage{titlesec}
\usepackage[colorlinks,bookmarks=false,citecolor=blue,linkcolor=red,urlcolor=blue]{hyperref}
\usepackage{epsfig}
\usepackage{color}
\usepackage{subfigure}
\usepackage{graphicx} 
\usepackage{dcolumn}  
\usepackage{mwe}
\input{insbox}

\newcommand{\be}{\begin{equation}}
\newcommand{\ee}{\end{equation}}

\begin{document}

\title{Spin-reversal energy barriers of 305 K\\
       for Fe$^{2+}$ $d^{6}$ ions with linear ligand coordination}

\author{Lei Xu}
\affiliation{Institute for Theoretical Solid State Physics, IFW Dresden, Helmholtzstr.~20, 01069 Dresden, Germany}

\author{Ziba Zangeneh}
\affiliation{Institute for Theoretical Solid State Physics, IFW Dresden, Helmholtzstr.~20, 01069 Dresden, Germany}

\author{Ravi Yadav}
\affiliation{Institute for Theoretical Solid State Physics, IFW Dresden, Helmholtzstr.~20, 01069 Dresden, Germany}

\author{Stanislav Avdoshenko}
\affiliation{Institute for Solid State Research, IFW Dresden, Helmholtzstr.~20, 01069 Dresden, Germany}

\author{Jeroen van den Brink}
\affiliation{Institute for Theoretical Solid State Physics, IFW Dresden, Helmholtzstr.~20, 01069 Dresden, Germany}
\affiliation{Department of Physics, Technical University Dresden, Helmholtzstr.~10, 01069 Dresden, Germany}

\author{Anton Jesche}
\affiliation{Center for Electronic Correlations and Magnetism, Augsburg University, Universit\"atsstr.~1, 86135 Augsburg, Germany}

\author{Liviu Hozoi}
\affiliation{Institute for Theoretical Solid State Physics, IFW Dresden, Helmholtzstr.~20, 01069 Dresden, Germany}

\begin{abstract}
A remarkably large magnetic anisotropy energy of 305 K is computed by quantum chemistry methods for
divalent Fe$^{2+}$ $d^6$ substitutes at Li-ion sites with $D_{6h}$ point-group symmetry within the
solid-state matrix of Li$_3$N.
This is similar to values calculated by the same approach and confirmed experimentally for linearly
coordinated monovalent Fe$^{1+}$ $d^7$ species, among the largest so far in the research area of 
single-molecule magnets.
Our {\it ab initio} results therefore mark a new exciting exploration path in the search for superior
single-molecule magnets, rooted in the $d_{xy}^{1.5} d_{x^2-y^2}^{1.5} d_{z^2}^1 d_{yz}^1 d_{zx}^1$
configuration of $d^6$ transition-metal ions with linear or quasilinear nearest-neighbor coordination.
This $d^6$ axial anisotropy may be kept robust even for symmetries lower than $D_{6h}$, provided the
ligand and farther-neighbor environment is engineered such that the
$d_{xy}^{1.5} d_{x^2-y^2}^{1.5} d_{z^2}^1 d_{yz}^1 d_{zx}^1$\,--\,$d_{xy}^{1} d_{x^2-y^2}^{1} d_{z^2}^2 d_{yz}^1 d_{zx}^1$
splitting remains large enough.
\end{abstract}

\date\today
\maketitle

The notion of single-molecule magnet (SMM) came into the field of quantum magnetism with recognizing
that certain molecules may display, as individual entities, the essential features of magnetic
nanoparticles:
large-spin electron configurations, strong axial anisotropy and a sizable energy barrier between
the two stable orientations of the total magnetic moment such that, below a certain `blocking'
temperature, the system can be trapped in one of those two states. 
%
  Below the blocking temperature SMM's exhibit therefore magnetic hysteresis.
Such effects have only been observed at rather low temperatures so far but intensive work is going
on to identify systems with superior properties in this regard: higher blocking temperatures,
longer relaxation times and larger coercivity fields.
An impelling idea is realizing regular, stable arrays of such molecules for high-density data
storage \cite{fe_d7_Bogani_2008}, provided that the associated blocking temperatures and
relaxation times are appropriately optimized.

SMM physics was first pointed out by Sessoli {\it et al.}~for a Mn$_{12}$ complex, in 1993
\cite{fe_d7_sessoli_1993}.
Since then the field advanced dramatically, with dozens of new SMM's being reported, either
$d$-metal or $f$-metal based.
As concerns their specific magnetic properties, the most remarkable are nowadays the Tb$^{3+}$
and Dy$^{3+}$ SMM's with N$_2^{3-}$ ligand bridges \cite{fe_d7_Rinehart_2011,fe_d7_Rinehart2_2011},
some lanthanide single-ion magnets with high-symmetry environment \cite{fe_d7_Gupta_2016}, the
fullerene-encapsulated $f$-electron SMM's \cite{fe_d7_Svitova_2014} and the linear Fe$^{1+}$
complexes \cite{fe_d7_Zadrozny_2013}.
Interestingly, SMM-like behaviour has been also identified recently for linearly coordinated Fe-ion
substitutes within the solid-state matrix of Li$_{3}$N \cite{fe_d7_Jesche_2014}.
The latter findings \cite{fe_d7_Zadrozny_2013,fe_d7_Jesche_2014} open new research avenues in this
field because, due to the well known `orbital quenching' issue in transition-metal (TM) compounds,
mononuclear $d$-metal ions have been rarely considered as good candidates to achieving first rate SMM 
characteristics.

The electronic structure and magnetic anisotropy of Fe ions placed within the Li$_{3}$N lattice
have been investigated on the theoretical side by calculations based on density functional theory
(DFT) \cite{fe_d7_Klatyk_2002,fe_d7_Novak_2002,fe_d7_Antropov_2014,fe_d7_Ke_2015}.
A Fe$^{1+}$ $d^7$ valence electron configuration has been assumed in the DFT studies
\cite{fe_d7_Novak_2002,fe_d7_Antropov_2014,fe_d7_Ke_2015} but diffraction \cite{Gregory_2002,Gordon_2004}
and x-ray absorption experiments on TM centers within the Li$_{3}$N matrix suggest 2+ valence states
for $d$-metal ions in such an environment \cite{fe_d7_Muller-Bouvet_2014}.
Here we provide unbiased {\it ab initio} results of many-body quantum chemistry calculations for
both Fe$^{1+}$ $d^7$ and Fe$^{2+}$ $d^6$ species at a Li lattice site.
The computed Fe$^{1+}$ $d^7$ excitation spectrum indicates an axial magnetic anisotropy of 31 meV for
linear N-Fe-N coordination, in agreement with experimental results for relatively large amount of
Fe cation substitution \cite{fe_d7_Jesche_2014,fe_d7_Jesche_2015}.
What is more, the calculated magnetic anisotropy reaches values of similar magnitude for Fe$^{2+}$
$d^6$, 26.3 meV (i.e., 305 K), if the overall lattice symmetry is not broken by vacancies in the immediate
neighborhood.
This is related to an unexpected $d_{z^2}^1 d_{xy}^{1.5} d_{x^2-y^2}^{1.5} d_{yz}^1 d_{zx}^1$
ground-state configuration in which due to subtle many-body effects one electron is removed from
the `deeper' $d_{z^2}$ level \cite{fe_d7_Novak_2002,fe_d7_Antropov_2014,fe_d7_Ke_2015}, as compared
to the Fe$^{1+}$ $d_{z^2}^2 d_{xy}^{1.5} d_{x^2-y^2}^{1.5} d_{yz}^1 d_{zx}^1$ ground state.
With a vacant nearest-neighbor (NN) Li site --- which ensures charge neutrality and coincides with
a Li$_{3-2x}$\,TM$_x^{2+}$N picture \cite{fe_d7_Muller-Bouvet_2014} for the substitution process ---
the interaction between the
$d_{z^2}^1 d_{xy}^{1.5} d_{x^2-y^2}^{1.5} d_{yz}^1 d_{zx}^1$ and
$d_{z^2}^2 d_{xy}^{1}   d_{x^2-y^2}^{1}   d_{yz}^1 d_{zx}^1$
states, arising from breaking the symmetry around the Fe$^{2+}$ ion, reduces the magnetic
anisotropy to $\approx$15 meV.
The latter number provides an explanation for the strong reduction of the magnetic anisotropy observed
experimentally in the very dilute system \cite{fe_d7_Jesche_2015} and support for the
Li$_{3-2x}$\,TM$_x^{2+}$N model \cite{fe_d7_Muller-Bouvet_2014} at very small concentrations
of TM centers.
Corroborated with earlier experimental results \cite{fe_d7_Muller-Bouvet_2014,fe_d7_Jesche_2015},
our computational data can therefore reconcile the Li$_{3-x}$\,TM$_x^{1+}$N and
Li$_{3-2x}$\,TM$_x^{2+}$N cation-substitution models, suggesting that due to finite concentration
of Li-ion vacancies the TM 2+ valence state dominates in the very dilute TM:Li$_{3}$N system
while TM 1+ plays a dominant role at large concentrations of TM species.
Most importantly, our findings draw attention to the very large magnetic anisotropy associated
with the Fe$^{2+}$ $d_{z^2}^1 d_{xy}^{1.5} d_{x^2-y^2}^{1.5} d_{yz}^1 d_{zx}^1$ configuration
in full $D_{6h}$ symmetry, analogous to Fe$^{1+}$ values defining so far the largest magnetic
anisotropies in the SMM research area \cite{fe_d7_Zadrozny_2013,fe_d7_Jesche_2014}.

\begin{figure*}
 \centering
 \subfigure{
 \hspace{-0.5cm}
\raisebox{5.8cm}{\textbf{ \large a}} \hspace{0.4cm}
\includegraphics[width=0.42\textwidth]{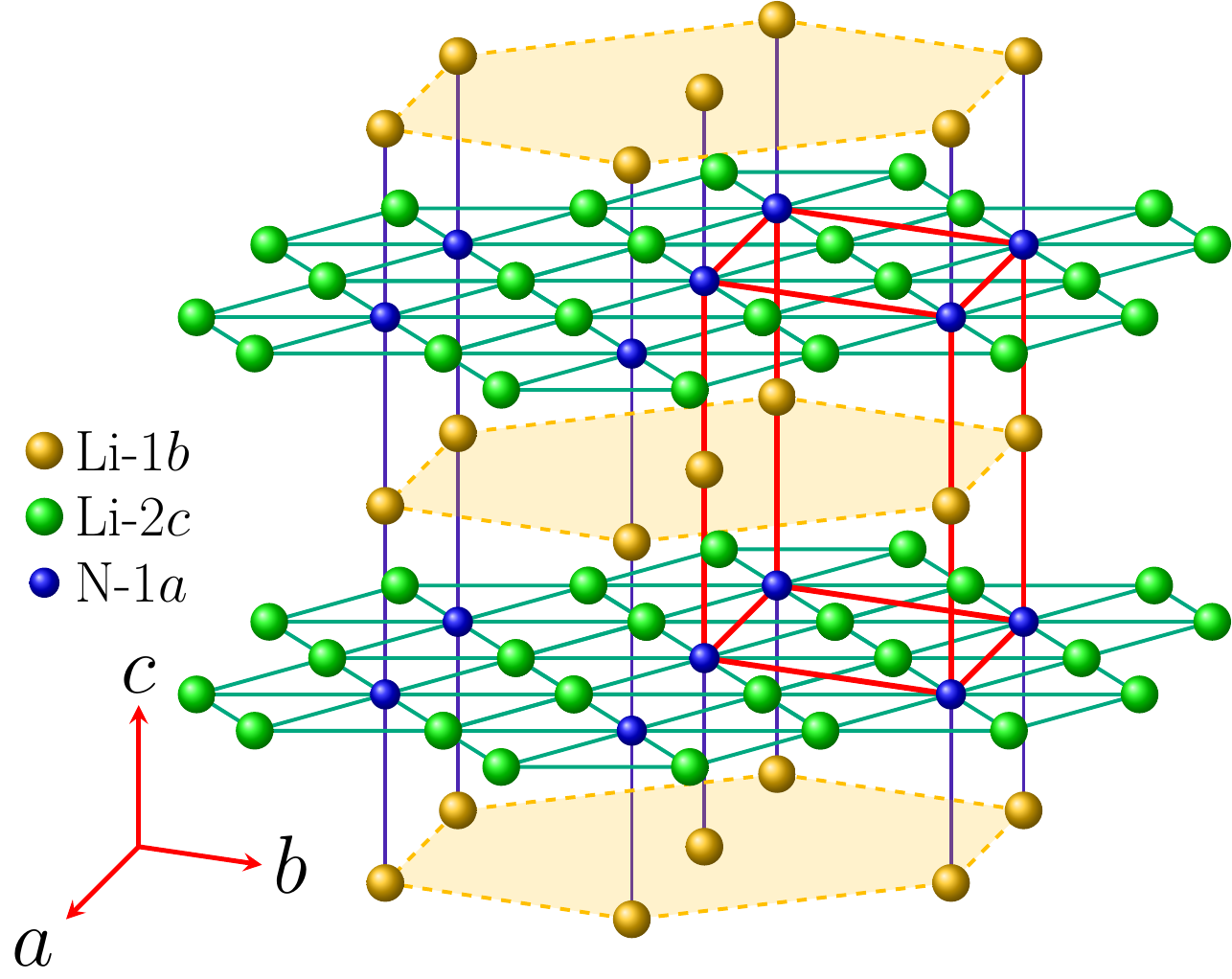}
\label{str_NLi3}}
 \quad
\subfigure{
\raisebox{5.8cm}{\textbf{\large b}}
\includegraphics[width=0.42\textwidth]{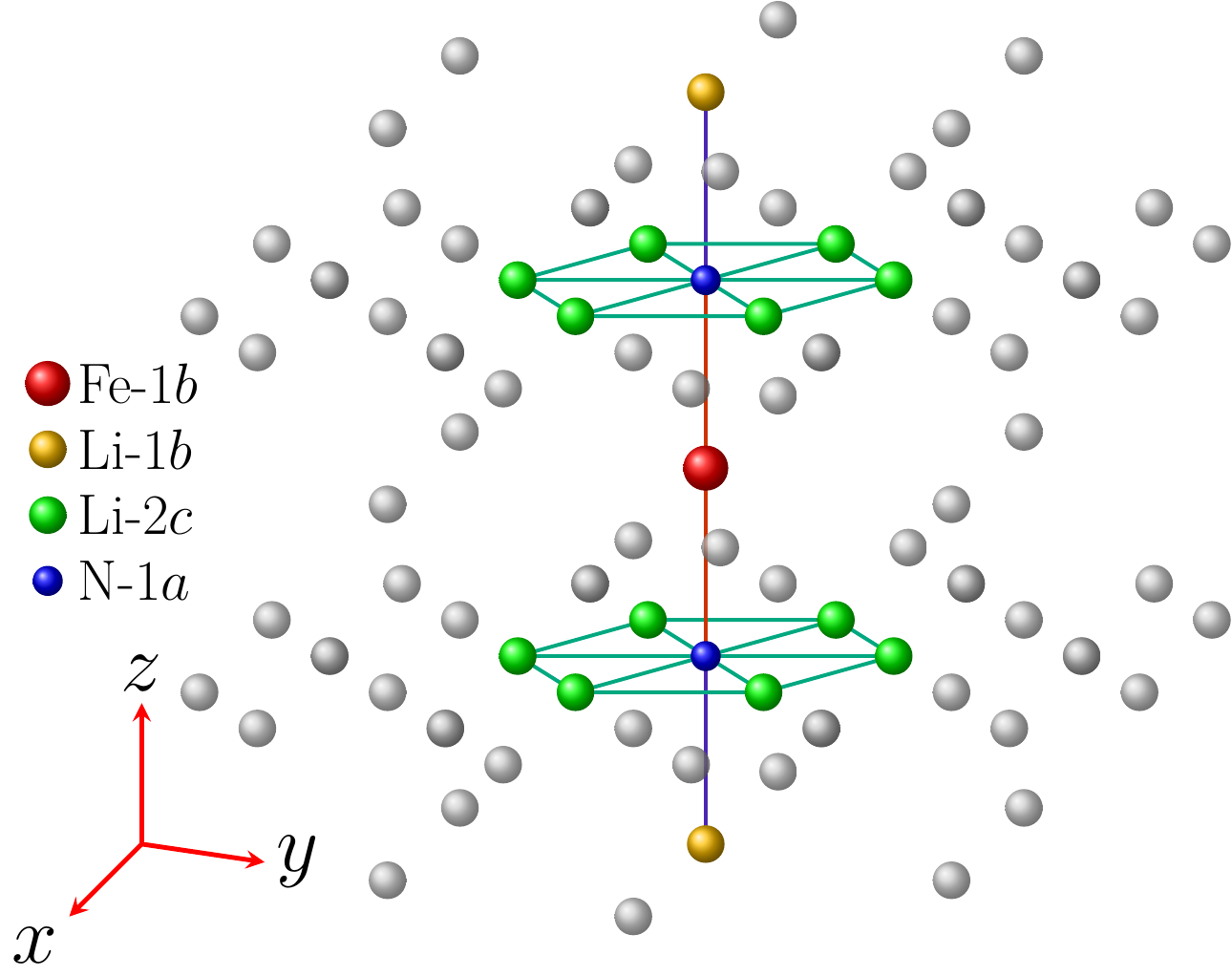}
\label{str_cluster}} \\
 \caption{
a) Crystal structure of Li$_{3}$N, with Li$_{2}$N honeycomb layers separated by Li-$1b$ sites.
The crystallographic unit cell is indicated as a red polyhedron.
b) Configuration of nearby sites around a Fe cation at the $1b$ crystallographic position in Li$_3$N.
These sites define the fragment treated at the all-electron quantum mechanical level in our calculations.
The extended solid-state surroundings are modeled as a large array of point charges,
depicted here as small grey spheres.
A similar type of linear coordination of the Fe ion (Fe$^{1+}$ $d^{7}$) is found in certain molecular
systems \cite{fe_d7_Zadrozny_2013}.
}
\end{figure*}

Relative energies describing the excitation spectrum of the Fe$^{1+}$ 3$d^7$ center are provided
in Table\;\ref{Fe_d7}.
We focus on cation substitution at linearly coordinated $1b$ Li sites, since that is the
geometrical configuration maximizing the single-ion magnetic anisotropy
\cite{fe_d7_Novak_2002,fe_d7_Zadrozny_2013}.
The many-body quantum chemistry calculations were performed on [FeN$_2$Li$_{14}$]$^{9+}$ clusters
as depicted in Fig.\,1, embedded within a large array of point charges which reproduces the Madelung
field of the Li$_3$N lattice.
All-electron basis sets as described in Methods were employed for the [FeN$_2$Li$_{14}$] unit.
In a first step, the orbitals were optimized for an average of all $d^7$ high-spin ($S\!=\!3/2$)
states using the multiconfigurational complete-active-space self-consistent-field (CASSCF)
approach \cite{QC_helgaker_2000}.
This ensures a balanced description of all $d^7$ electron configurations:
$a_{1g}^2 e_{2g}^3 e_{1g}^2$,
$a_{1g}^2 e_{2g}^2 e_{1g}^3$,
$a_{1g}^1 e_{2g}^3 e_{1g}^3$,
$a_{1g}^1 e_{2g}^4 e_{1g}^2$ and
$a_{1g}^1 e_{2g}^2 e_{1g}^4$,
where $d_{z^2}$ belongs to the $A_{1g}$ irreducible representation, $\{d_{xy},d_{x^2-y^2}\}$ to
$E_{2g}$ and $\{d_{yz},d_{zx}\}$ to $E_{1g}$, for $D_{6h}$ point-group symmetry.
Following the CASSCF calculation, multireference configuration-interaction (MRCI) computations
with single and double excitations were performed \cite{QC_helgaker_2000,mrci_werner_88}.
Spin-orbit couplings were subsequently accounted for according to the procedure described in
Ref.\cite{SOC_molpro}.
We utilized the quantum chemistry package {\sc molpro} \cite{molpro12} and Li$_3$N lattice parameters as
derived in Ref.\cite{Huq_2007}.
Yet we allowed relaxation of the N-Fe-N bonds, i.e., for the two nitrogen ions adjacent to the Fe
cation we determined the $z$-axis positions which minimize the total energy while fixing all other
lattice coordinates as in the unmingled Li$_3$N crystal.
At the MRCI level, the `relaxed' Fe$^{1+}$-N bond lengths are 1.92 \AA , slightly shorter than 
the experimental Li-N distances along the $z$ axis \cite{Huq_2007}.

In the described MRCI+SOC computational frame, we predict three main sets of excited states
(see Table\;\ref{Fe_d7}):
low-lying excited states related to the magnetic anisotropy of the ($S\!=\!3/2$, $L\!=\!2$)
$a_{1g}^2 e_{2g}^3 e_{1g}^2$ configuration in the range of $\lesssim$100 meV, 
high-spin $e_{2g}$ to $e_{1g}$ excitations at 1.1--1.2 eV and
a multitude of crystal-field excitations from 1.75 eV onwards.
The $a$ and $b$ labels in Table\;\ref{Fe_d7} are used in order to distinguish between states implying
the same electron configuration, irreducible representation and spin multiplicity.
The spin-orbit treatment was carried out in terms of all $S\!=\!3/2$ quartets and those doublets
with MRCI relative energies of less than 2.5 eV.
The lowest excited state, defining the magnetic anisotropy energy, lies in this case at 30 meV.
If the orbitals are optimized just for the $a_{1g}^2 e_{2g}^3 e_{1g}^2$ ground-state configuration,
this particular excitation energy changes to 31 meV.

\begin{table}[t]
\caption{
$3d$-shell energy levels for a Fe$^{1+}$ ion at the Li $1b$ crystallographic position in
Li$_3$N; unless otherwise specified, units of eV are used.
All $3d^7$ $S\!=\!3/2$ and the few lowest $S\!=\!1/2$ states are listed.
The spin-orbit calculations provide three main groups of Kramers doublets:
between 0 and 100 meV, at 1.1--1.2 eV and from 1.75 eV onwards.
}
\begin{tabular}{llll}
\hline
\hline\\[-0.30cm]
Fe$^{1+}$ $3d^7$ splittings                      &CASSCF  &MRCI   &MRCI+SOC\\
\hline
\\[-0.20cm]
$   ^4\!E_{2g}$ ($a_{1g}^2 e_{2g}^3 e_{1g}^2$)   &0       &0      &0, 30, 62, 96 meV  \\
$   ^4\!E_{1g}$ ($a_{1g}^2 e_{2g}^2 e_{1g}^3$)   &0.91    &1.09   &1.11\,--\,1.16     \\[0.20cm]

$a\,^4\!E_{1g}$ ($a_{1g}^1 e_{2g}^3 e_{1g}^3$)   &1.50    &1.78   &1.75              \\
$a\,^4\!A_{2g}$ ($a_{1g}^1 e_{2g}^4 e_{1g}^2$,  
                 $a_{1g}^1 e_{2g}^2 e_{1g}^4$)   &1.67    &1.87   &$|$                \\

$a\,^2\!E_{2g}$ ($a_{1g}^2 e_{2g}^3 e_{1g}^2$)   &2.23    &2.05   &$|$                \\
$   ^2\!E_{1g}$ ($a_{1g}^2 e_{2g}^4 e_{1g}^1$)   &2.25    &2.11   &$|$              \\
$b\,^2\!E_{2g}$ ($a_{1g}^2 e_{2g}^3 e_{1g}^2$)   &2.29    &2.11   &$|$              \\
$   ^2\!E_{1g}$ ($a_{1g}^2 e_{2g}^3 e_{1g}^2$)   &2.50    &2.35   &2.41               \\[0.20cm]

$b\,^4\!E_{1g}$ ($a_{1g}^1 e_{2g}^3 e_{1g}^3$)   &2.69    &2.70   &       \\ 
$b\,^4\!A_{2g}$ ($a_{1g}^1 e_{2g}^4 e_{1g}^2$,  
                 $a_{1g}^1 e_{2g}^2 e_{1g}^4$)   &3.27    &3.28   &             \\[0.06cm]
\hline
\hline
\end{tabular}
\label{Fe_d7}
\end{table}

Results of similar type are provided in Table\;II for the Fe$^{2+}$ $d^6$ configuration.
What makes this valence electron configuration worth investigating is the observation that a 
finite amount of vacant Li sites would necesarilly require a higher ionized state for some of
the TM centers, according to a Li$_{3-x-2y}$Fe$_{x}^{1+}$Fe$_{y}^{2+}$N picture.
The very surprising result for the Fe$^{2+}$ $d^6$ ion is that the computed ground-state electron
configuration defies a simple diagram of single-electron levels according to which, from the
Fe$^{1+}$ $a_{1g}^2e_{2g}^3e_{1g}^2$ `reference', removal of one additional electron yields a
          $a_{1g}^2e_{2g}^2e_{1g}^2$ orbital occupation \cite{fe_d7_Antropov_2014,fe_d7_Ke_2015}.
Instead, the quantum chemistry calculations indicate the 3$d$-shell Coulomb interactions are
such that it is energetically more favorable to remove one electron from the apical $a_{1g}$
$d_{z^2}$ orbital rather than further depleting the `in-plane' $e_{2g}$'s, $d_{xy}$ and 
$d_{x^2-y^2}$.
An important detail here is that there are no negatively charged ions in the plane within which
the lobes of the latter lie while the former points to anions with formal $3-$ charges.
Consequently, the 3$d^6$ ground-state configuration is $a_{1g}^1 e_{2g}^3 e_{1g}^2$ according
to our calculations, with an occupation of the $e_g$ levels that provides again a large angular
momentum ($L\!=\!2$) and strong axial anisotropy.
%
%
%
%
Using orbitals optimized for an average of all $S\!=\!2$ $d^6$ states, the magnetic anisotropy
energy comes as 26.3 meV in the $d^6$ spin-orbit MRCI calculation (see Table\;II);
the same value, 26.3 meV, is obtained with orbitals optimized just for the lowest two quintet
states.
That is 305 K, room-temperature energy scale.

Also for these computations, we considered all the high-spin ($S\!=\!2$) states in the spin-orbit
treatment but only the spin triplets and singlets with MRCI relative energies of less than 2.8 eV.
As for the Fe$^{1+}$ $d^7$ ion, the different $m_{\mathrm J}$ states associated with the ground-state
configuration cover an energy window extending up to $\approx$100 meV.
The first crystal-field excitation, however, implies here an energy scale of only $\approx$200 meV;
that is the $^5\!E_{2g}$ ($a_{1g}^1e_{2g}^3e_{1g}^2$) to $^5\!A_{1g}$ ($a_{1g}^2 e_{2g}^2 e_{1g}^2$)
transition.
Other excited states lie in the energy range of 1.2--1.3 eV and from 1.8 eV onwards.
All these results correspond to `relaxed' Fe$^{2+}$-N bonds of 1.88 \AA .
Significant shortening of the N-TM-N bonds has been also inferred from EXAFS measurements on TM ions
embedded within the solid-state Li$_{3}$N matrix \cite{fe_d7_Muller-Bouvet_2014}.

\begin{table}[t]
\caption{
$3d$-shell energy levels for a Fe$^{2+}$ ion at the Li $1b$ crystallographic position in
Li$_3$N; unless otherwise specified, units of eV are used.
All $3d^6$ $S\!=\!2$ and the lowest $S\!=\!1$ and $S\!=\!0$ states are listed.
The spin-orbit calculations provide four main groups of excited states:
up to $\approx$105 meV, 180--200 meV, 1.2--1.3 eV and from 1.8 eV onwards.
}
\begin{tabular}{llll}
\hline
\hline\\[-0.30cm]
Fe$^{2+}$ $3d^6$ splittings                      &CASSCF  &MRCI   &MRCI+SOC\\
\hline
\\[-0.20cm]
$   ^5\!E_{2g}$ ($a_{1g}^1 e_{2g}^3 e_{1g}^2$)   &0       &0      &0, 26, 52, 78, 104 meV  \\ 
$   ^5\!A_{1g}$ ($a_{1g}^2 e_{2g}^2 e_{1g}^2$)   &0.26    &0.14   &0.18\,--\,0.19     \\
$   ^5\!E_{1g}$ ($a_{1g}^1 e_{2g}^2 e_{1g}^3$)   &1.09    &1.18   &1.21\,--\,1.26     \\[0.35cm]

$a\,^3\!E_{1g}$ ($a_{1g}^2 e_{2g}^3 e_{1g}^1$)   &2.31    &1.84   &1.84     \\
$b\,^3\!E_{1g}$ ($a_{1g}^2 e_{2g}^3 e_{1g}^1$)   &2.37    &1.89   &$|$     \\

$a\,^3\!E_{2g}$ ($a_{1g}^1 e_{2g}^3 e_{1g}^2$)   &2.58    &2.38   &$|$     \\
$b\,^3\!E_{2g}$ ($a_{1g}^1 e_{2g}^3 e_{1g}^2$)   &2.74    &2.55   &$|$     \\
$   ^3\!A_{2g}$ ($a_{1g}^1 e_{2g}^3 e_{1g}^2$)   &2.81    &2.55   &$|$    \\
$   ^1\!A_{1g}$ ($a_{1g}^2 e_{2g}^4$ )           &3.34    &2.77   &2.83               \\[0.06cm] 
\hline
\hline
\end{tabular}
\label{Fe_d6}
\end{table}

To retain overall charge neutrality, for the set of calculations whose results are summarized in
Table\;II, we compensated the larger, 2+ valence state of the Fe ion by adding one (negative)
electronic charge to the nearby crystalline surroundings.
In particular, we equally distributed this elementary negative charge over the six closest Li 1$b$
sites within the $xy$ plane.
This way, the $D_{6h}$ point-group symmetry at the Fe site is preserved.
An additional set of calculations was then performed with one of the NN Li ions at a 2$c$ 
crystallographic position explicitly removed from the cluster described by quantum chemistry
methods.
This also preserves overall charge neutrality, according to the Li$_{3-2x}$\,TM$_x^{2+}$N model
of Muller-Bouvet {\it et al.}~\cite{fe_d7_Muller-Bouvet_2014}.
The symmetry being lower with such a Li vacant site, the $a_{1g}^1 e_{2g}^3 e_{1g}^2$ and
$a_{1g}^2 e_{2g}^2 e_{1g}^2$ configurations, in particular, can interact and admix.
As a result, the low-energy part of the MRCI spectrum displays now a richer structure:
with orbitals optimized in the prior CASSCF step just for the $a_{1g}^1e_{2g}^3e_{1g}^2$
and $a_{1g}^2 e_{2g}^2 e_{1g}^2$ configurations and maximum spin multiplicity, the relative
energies of the spin-orbit MRCI states are 0 (two states), 15, 17, 23, 81, 84, 86 and 94
(again as a doublet) meV.
In other words, the magnetic anisotropy energy of the Fe$^{2+}$ $d^6$ ion is reduced from
$\Delta^{d^6}_{D_{6h}}\!=\!26$ meV in $D_{6h}$ symmetry to
$\Delta^{d^6}_{C_{s }}\!=\!15$ meV when the symmetry is broken by creating a vacancy at a NN Li
site.
Analysis of the spin-orbit wave functions shows that these effects imply admixture(s) of the
$a_{1g}^2 e_{2g}^2 e_{1g}^2$ components of only tenths of 1\% to the low-lying
$a_{1g}^1 e_{2g}^3 e_{1g}^2$ states.
Obviously, the more complicated structure of the spectrum in the lower-symmetry case is related to
having slightly different degrees of $a_{1g}^1 e_{2g}^3 e_{1g}^2$\,--\,$a_{1g}^2 e_{2g}^2 e_{1g}^2$
admixture for different spin-orbit eigenvectors.
The relaxed Fe-N bond lengths are in both cases, without and with a Li-ion vacancy, 1.88 \AA .
According to experimental investigations \cite{Gregory_2002,Gordon_2004}, the  Li-ion vacancies mainly occur
within the Li$_2$N planes.

Remarkably, our computational results for the magnetic anisotropies of Fe$^{1+}$ $d^7$ and Fe$^{2+}$
$d^6$ centers within the solid-state matrix of Li$_3$N find strong support in recent experimental
data on Li$_{3-x-2y}$Fe$_{x}^{1+}$Fe$_{y}^{2+}$N,
that indicate magnetic anisotropy energies $\Delta_{x\to 0}\!=\!13$ meV in the very dilute case
and $\Delta_{x\gg y}\!=\!27$ meV for high concentration of Fe \cite{fe_d7_Jesche_2015}.
These drastic variations in the measurements can be assigned to the presence of a finite amount of
Li-ion vacancies.
An intrinsic load of vacant Li sites has been indeed found experimentally for Li$_3$N, in the
range of $\sim$1\% \cite{schulz_defect_1979}, which suggests that for keeping overall charge
neutrality Fe ions in the immediate neighborhood of such vacancies might adopt a Fe$^{2+}$ $d^6$
configuration.
%
%
The reason no connection has been made so far between these variations of the magnetic properties
and the possible predominance of Fe$^{2+}$ $d^6$ species in the very dilute case is the fact that 
the $d^6$ ground-state configuration is ussually associated for linear coordination with a
$d_{z^2}^2d_{xy}^1d_{x^2-y^2}^1d_{yz}^1d_{zx}^1$ orbital occupation \cite{fe_d7_Novak_2002,fe_d7_Antropov_2014,fe_d7_Ke_2015}
for which single-ion anisotropy can only occur through weaker, second-order SOC's.
The good agreement between our MRCI value $\Delta^{d^6}_{C_{s }}\!=\!15$ meV and the experimentally
derived $\Delta_{x\to 0}\!=\!13$ meV makes therefore plausible the scenario in which the magnetic properties
in the dilute regime are mainly determined by Fe$^{2+}$ $d^6$ ($a_{1g}^1 e_{2g}^3 e_{1g}^2$) ions
with broken-symmetry nearby surroundings.
Additional support is provided by the good agreement between the MRCI result $\Delta^{d^7}_{D_{6h}}\!=\!30$ 
meV (see Table\;\ref{Fe_d7}) and the experimental estimate $\Delta_{x\gg y}\!=\!27$ meV at large
concentrations of Fe.

Resonant inelastic x-ray scattering (RIXS) measurements on the Fe:Li$_3$N system might throw fresh
light on the problem.
The high resolution achieved nowadays in RIXS should allow to directly verify our prediction of a 
distinct peak at 0.15--0.20 eV for the Fe$^{2+}$ $d^6$ electron configuration.
According to our computational results, the position of these crystal-field excited states is about
the same in $D_{6h}$ symmetry (see Table\;II) and when the symmetry is broken by creating a Li
vacancy next to the Fe$^{2+}$ $d^{6}$ center.
That post-CASSCF quantum chemistry calculations can describe the RIXS $d$--$d$ excitation spectra
with very good accuracy has been convincingly shown already for TM ions in a variety of environments
\cite{Minola_2013,rixs_raspt2_guo16,rixs_raspt2_kunnus16,Nikolay_RIXS}.
Another experimental technique capable of verifying the existence of Fe$^{2+}$ $d^6$ ions in
Fe:Li$_3$N is M\"{o}ssbauer spectroscopy.
In addition to stimulating further experimental investigations, our computational data define the
frame for subsequent model-Hamiltonian constructions for addressing the magnetodynamics of this
system \cite{Lunghi_2017,Layfield_2017}.
Aspects which remain to be clarified is the role of spin-phonon couplings in under-barrier spin
relaxation \cite{Lunghi_2017} but also the occurence of clustering effects among the Fe-ion substitutes and of
sizable magnetic exchange between proximate Fe sites.

To summarize, our {\it ab initio} data put into the spotlight the linearly coordinated Fe$^{2+}$ $d^6$
ion as candidate for viable SMM behaviour.
The calculated magnetic anisotropy splitting of 26.3 meV (i.e., 305 K) in $D_{6h}$ symmetry compares
favorably to values measured (28 meV \cite{fe_d7_Zadrozny_2013} and 27--37 meV \cite{fe_d7_Jesche_2014,fe_d7_Jesche_2015})
or computed by similar theoretical methods (26 meV \cite{fe_d7_Zadrozny_2013}) for Fe$^{1+}$ $d^7$
species with linear coordination, among the largest so far in the research area of SMM's.
This substantial spin-reversal energy barrier of the Fe$^{2+}$ ion is associated with a $a_{1g}^1 e_{2g}^3 e_{1g}^2$
ground-state electron configuration, not anticipated by earlier DFT calculations for TM species in such 
an environment \cite{fe_d7_Novak_2002,fe_d7_Antropov_2014,fe_d7_Ke_2015} and made possible through a
subtle interplay between ligand/crystal-field splittings and on-site Coulomb interactions.
The effects we point out here warrant more detailed investigations of both iron(I) and iron(II)
complexes with linear or quasilinear two-ligand coordination.
For iron(II), engineering of the $^5E_{2g}$\,--\,$^5\!A_{1g}$ splitting towards larger values would
allow large magnetic anisotropy barriers also for symmetries much lower than $D_{6h}$.

\noindent
\vspace{5mm}

\noindent
{\bf {\large Methods}}

\noindent
All {\it ab initio} calculations were carried out with the quantum chemistry package {\sc molpro}
\cite{molpro12}, using the room-temperature lattice parameters reported in Ref.\,\cite{Huq_2007}.
To compute the magnetic anisotropy and the on-site $d$--$d$ excitation spectrum, an embedded
cluster consisting of one central Fe ion, the two NN NLi$_{6}$ hexagonal plaquettes and the nearby
two Li sites on the $z$ axis was considered.
The solid-state surroundings were modeled as a finite array of point charges fitted \cite{Ewald_soft}
to reproduce the crystal Madelung field in the cluster region \cite{madpot_roos}.
We applied all-electron Douglas-Kroll basis sets of triple-zeta quality for the central Fe ion
\cite{Sc-Zn} and all-electron triple-zeta basis sets for the two NN nitrogen ligands \cite{Li-Ne}
and the Li species, supplemented with polarization functions. 
For the CASSCF calculations, we employed an active space of five $3d$ orbitals at the Fe
site and seven (six) electrons for the Fe$^{1+}$ $d^{7}$ (Fe$^{2+}$ $d^{6}$) valence configuration;
the orbitals were optimized for an average of all high-spin (either $S$=3/2 or $S$=2)
states.
Only the N $2s,2p$ and Fe $3s,3p,3d$ electrons were correlated in the subsequent MRCI treatment.
\\

\vspace{3mm}
\noindent
\noindent
{\bf {\large Acknowledgements}}

\noindent
Calculations were performed at the High Performance Computing Center (ZIH) of the Technical
University Dresden (TUD).
L.\,X., Z.\,Z., R.\,Y., S.\,A. and L.\,H.~thank U.~Nitzsche for technical assistance.
We acknowledge financial support from the German Science Foundation (Deutsche Forschungsgemeinschaft,
DFG --- HO-4427/2, JE-748/1 and SFB-1143) and thank D. Efremov and A. Popov for instructive discussions.

\bibliography{biblio_jun21}
\bibliographystyle{rsc} 

\end{document}